\documentclass{ws-ijmpa}

\usepackage[super]{cite}
\usepackage{combelow}
\usepackage{xcolor}

\usepackage[verbose,hypertexnames=false]{hyperref}
\hypersetup{colorlinks=false,allbordercolors=blue,pdfborderstyle={/S/U/W 1}}

\allowdisplaybreaks

\usepackage{systeme}
\usepackage{mathtools}
\usepackage[inline]{enumitem}
\usepackage{graphicx}

\usepackage{braket}
\usepackage{nicefrac}
\usepackage{bm}
\usepackage{bbm}
\usepackage{braket}
\usepackage{bbold}    
\usepackage{slashed}
\usepackage[normalem]{ulem}
\usepackage{array}

\newcommand{\beqn}{\begin{eqnarray}}
\newcommand{\eeqn}{\end{eqnarray}}
\newcommand{\beqs}{\begin{subequations}}
\newcommand{\eeqs}{\end{subequations}\\[-2mm]\noindent}

\newcommand{\bs}{\boldsymbol}
\newcommand{\avr}[1]{{\left\langle #1 \right\rangle}}

\begin{document}

\title{On conformal anomaly and pair production}

\author{M. N. Chernodub}
\address{Institut Denis Poisson CNRS UMR 7013, Universit\'e de Tours, 37200 Tours, France\\
Department of Physics, West University of Timi\cb{s}oara,  Bd.~Vasile P\^arvan 4, Timi\cb{s}oara 300223, Romania}

\maketitle

\begin{abstract}
We conjecture that the time-reversal-even component of the pair production rate $\Gamma$ of particles in background fields in conformal 3+1 dimensional field theories is given by the anomalous trace $\langle T^\mu_{\ \mu}\rangle$ of the energy-momentum tensor: $\Gamma = (\pi/2) \langle T^\mu_{\ \mu} \rangle \Theta( \langle T^\mu_{\ \mu}\rangle)$, where $\Theta(x)$ is the Heaviside step function. We show that this relation, written in mostly-plus metric, correctly describes the one-loop Schwinger pair creation of massless particles both in scalar and spinor versions of quantum electrodynamics. It also accurately points to the Savvidi instability of the gluonic vacuum towards the formation of the chromomagnetic condensate. The conjectured formula also reproduces (presumably, non-Hawking) radiation generated by static gravitational fields in the absence of an event horizon via a new evaporation mechanism suggested in Ref.~[\citeonline{Wondrak2023}]. A concurrent mechanism of the particle production due to the axial anomaly is briefly discussed and critical remarks on the state-dependence are also given.
\end{abstract}


\ccode{PACS numbers: 04.60.-m, 03.70.+k, 11.10.-z}

\section{Introduction}

Signatures of vacuum instability in a strong electric field were first found in work by Sauter on the Klein paradox~\cite{Sauter:1931zz}. This fascinating effect has been recognized and developed further by Heisenberg and Euler~\cite{Heisenberg:1936nmg} and later formalized in terms of a pair production process in QED by Schwinger~\cite{Schwinger1951,Schwinger:1954zza}. The physical interpretation of this phenomenon, often called the Schwinger effect, is linked to the quantum fluctuations in which virtual pairs of electrons $e^-$ and positrons $e^+$ are constantly created in the vacuum to be annihilated shortly later according to the Heisenberg uncertainty principle. In a strong background electric field, the created $e^+ e^-$ particles are forced to flow away in opposite directions, they become spatially separated, and, without possibility to annihilate, become real particles. 

A similar process exists in gravitational fields near black holes. A black hole produces the Hawking radiation~\cite{Hawking1974,Hawking1975}, which can be associated with the particle tunneling process~\cite{Parikh2000} in which one particle from the pair, created in the vicinity of the event horizon, gets swallowed by the hole while another particle has sufficient energy to escape to infinity. The escaping particles form the outgoing energy flux, which diminishes the mass of the black hole and, therefore, leads to the black hole evaporation. Due to a nonlocality of the tunneling process, this effect operates in an extended vicinity above the black hole event horizon, thus creating the notion of the quantum atmosphere~\cite{Giddings:2015uzr} (see also \cite{Eune:2015xvx,Eune:2017iab}). It was recently realized that such quantum atmospheres could possess nontrivial thermodynamic features that can be probed in condensed matter experiments~\cite{Bermond:2022mjo}. 

In 1+1 spacetime dimensions, the Hawking radiation can be related to a gravitational (Einstein) anomaly, which implies a non-conservation of energy-momentum of a chiral particle in a curved spacetime~\cite{Robinson:2005pd}. The anomaly appears due to quantum fluctuations when classical symmetries are inconsistent with the quantization procedure~\cite{Bertlmann2000}. In 1+1  and 3+1 spacetimes, the Hawking effect can also be interpreted~\cite{Christensen:1977jc} in terms of conformal (or trace, related to scale~\cite{Nakayama:2013is}) anomalies~\cite{Duff:1993wm, Capper:1973mv, Capper:1974ed, Deser1976, Duff:1977ay}. See also Ref.~[\citeonline{Volovik:2023stf}] for discussions of different channels of radiation.

In addition, one can argue that the particle creation in a static gravitational field can also produce particles even without the event horizon~\cite{Wondrak2023}. In this scenario, the virtual pairs of particles are separated by local tidal forces and become real particles, similar to what happens in the Schwinger effect. Some of these real particles will fall to the gravitating body and will later be recaptured, while other particles will escape to infinity and create, similarly to the Hawking effect, an outgoing flux of matter~\cite{Wondrak2023}. 

We argue that in the off-event-horizon mechanism of Ref.~[\citeonline{Wondrak2023}] of particle pair production, the creation rate in the static, both gravitational and electric, background can be directly related to the conformal anomaly. This effect is different from the anomaly-assisted creation of particles in a dynamical gravitational field during the inflationary stage of the Universe evolution~\cite{Dolgov:1981nw,Dolgov:1993vg}.

\section{Particle production and effective action} 

The rate density of particle production events $d N/d t = \Gamma$ is determined by the imaginary part, $\Gamma = 2 \, {\mathrm{Im}}\, {\mathcal{L}}_{\mathrm{eff}}$, of the Lagrangian ${\mathcal{L}}_{\mathrm{eff}}$ associated with the effective action~\cite{Schwinger1951},
\begin{align}
    W = \int d^4 x \sqrt{-g} {\mathcal{L}}_{\mathrm{eff}}\,,
\label{eq_W_Leff}
\end{align}
which takes into account quantum effects. $\Gamma > 0$ implies non-persistence of vacuum due to pair creation~\cite{Schwinger1951,Schwinger:1954zza}. We set $\hbar = c = 1$ everywhere in the article and work in the mostly-plus metric convention $(-,+,+,+)$.

In our paper, we argue that in {\it conformal} field theories, the pair-production rate $\Gamma$ in background gravitational and gauge (electromagnetic or gluon) fields can be related to the conformal (trace) anomaly:
\begin{align}
    \Gamma = 
    \left\{
    \begin{array}{ll}
    \frac{\pi}{2} \avr{T^\mu_{\ \mu}}\,, &\qquad \avr{T^\mu_{\ \mu}} > 0\\[1mm]
    0\,, & \qquad {\rm otherwise}
    \end{array}
    \right.\,,
    \label{eq_main}
\end{align}
where $\avr{T^\mu_{\ \mu}} \equiv \avr{T^\mu_{\ \mu}}_{\mathrm{an}}$ is the anomalous trace of the energy-momentum tensor $T^{\mu\nu}$. Notice that Eq.~\eqref{eq_main} implies that the particle production rate $\Gamma$, contrary to the outgoing energy flux $T^{r}_{\ t}$, does not depend on the choice of quantum state due to the state independence of the trace anomaly (for a related discussion, see Refs.~\cite{Balbinot:1999ri, Balbinot:1999vg} as well as a comment below). 

Relation~\eqref{eq_main} has a quantum nature because, in classical conformal theories in an even number of spacetime dimensions, the trace of the stress-energy tensor vanishes identically, $(T^\mu_{\ \mu})_{\mathrm{cl}} \equiv 0$. Quantum fluctuations can violate this identity, $\avr{T^\mu_{\ \mu}} \neq 0$, hence the term ``trace anomaly'' or ``conformal anomaly''. 

In Eq.~\eqref{eq_main}, we used the convention that this equation has a relation to the pair production iff $\avr{T^\mu_{\ \mu}} > 0$. Otherwise, $\Gamma \leqslant 0$ is equivalent to $\Gamma \equiv 0$ because a negative production rate does not lead to the production of pairs.

The rate density of the particle production of $N$ massless scalar degrees of freedom (with $N=1$ for a neutral field and $N=2$ for a complex field) in the curved $d=3+1$ dimensional spacetime (described by the metric $g_{\mu\nu}$) in the presence of the classical electromagnetic field (characterized by the field strength $F_{\mu\nu}$) has been found recently~\cite{Wondrak2023}:
\begin{align} \label{eq_PRL_recent}
    \Gamma_{N\mathrm{sc}} = \frac{N}{32 \pi} & \biggl[ \frac{1}{180} 
    \bigl( R_{\mu\nu\alpha\beta} R^{\mu\nu\alpha\beta} - R_{\mu\nu} R^{\mu\nu} \bigr)  + \frac{1}{2} \left(\frac{1}{6} - \xi \right)^2 R^2 - \frac{q^2}{12} F_{\mu\nu} F^{\mu\nu} \biggr]\,,
\end{align}
where the curved background is expressed via the Riemann tensor $R_{\mu\nu\alpha\beta}$, the Ricci tensor $R_{\mu\nu} = R^{\alpha}_{\ \mu\alpha\nu}$, and the scalar curvature $R \equiv R^\mu_{\ \mu}$. 
In the case of a massless conformally coupled scalar field with $\xi = 1/6$, the gravitational part in the above formula agrees, up to a coefficient, with the results obtained by other methods~\cite{Frieman1989, Campos1994, Dobado1999, Akhmedov2024}. 

Since the right-hand side of Eq.~\eqref{eq_PRL_recent} involves only expressions that are invariant under the time-reversal ($T$) symmetry, this formula can be conjectured to capture only a $T$-even part of the particle production rate.~\cite{Ferreiro2024} And, indeed, the full production rate also involves a contribution from the axial anomaly, which necessarily involves $T$-odd scalars such as the second Lorentz invariant $F_{\mu\nu} {\tilde F}^{\mu\nu}$~\cite{Dunne:2004nc}. 

The quantity $q$ in Eq.~\eqref{eq_PRL_recent} is the electric charge of the scalar particle minimally coupled to electromagnetism. A neutral ($q=0$) scalar field carrying one degree of freedom ($N = 1$) is described by the following Lagrangian:
\begin{align}
    {\mathcal L} = - \frac{1}{2} \partial_\mu \phi \partial^\mu \phi - \frac{1}{2} \xi R \phi^2 - \frac{1}{2} m^2 \phi^2\,,
    \label{eq_L_scalar}
\end{align}
where $\xi$ couples the Ricci curvature scalar $R$ to the scalar field, with $\xi = 1/6$ being the conformal value. For consistency with previous studies, we also added the mass term, which will be set to zero later.

The remarkable feature of the pair-production effect~\eqref{eq_PRL_recent} is that it can take place in static gravitational fields, as opposed to dynamical gravitational fields~\cite{Parker:1968mv,Dolgov:1981nw,Dolgov:1993vg}, thus suggesting that this effect is a Hawking-type of radiation associated with the presence of an event horizon~\cite{Hawking1974,Hawking1975}. However, pair production~\eqref{eq_PRL_recent} takes place even in the absence of a horizon (for any massive object), which indicates that this phenomenon is either an addition or a generalization of Hawking radiation~\cite{Wondrak2023}. 

\section{Effective action, trace anomaly, and pair production}

We start from the simplest case of the scalar field for which the trace anomaly has been elaborated in great detail in Ref.~[\citeonline{Brown1977}]. Our relation~\eqref{eq_main} between the conformal (trace) anomaly and the off-event-horizon particle production can be deduced by matching the anomalous term in the one-loop effective action $W$ represented as an integral over the proper time $s$ of Ref.~[\citeonline{Brown1977}] with the representation of the same action in terms of the spectral parameter $s$ in the heat-kernel approach of Ref.~[\citeonline{Wondrak2023}] based on the Barvinsky--Vilkovisky expansion~\cite{Barvinsky:1990up}.

The one-loop action functional $W$ is given by a formal divergent expression $W = (i/2) \ln \det G^{-1}$, where $G(x,x') \equiv \avr{i T \bigl(\phi(x) \phi(x')\bigr)}$ represents the Green function associated with the quadratic Lagrangian~\eqref{eq_L_scalar}. The functional $W$  has a close relation to the expectation value of the energy-momentum tensor, $\avr{T^{\mu\nu}}$, in its response, $W \to W + \delta W$ to the metric variation, $g_{\mu\mu} \to g_{\mu\nu} + \delta g_{\mu\nu}$ in (even) $D$ spacetime dimensions:
\begin{align}
    \delta W = \frac{i}{2} {\mathrm{Tr}}\,G^* \delta G^{-1} = \int d^D x \, \sqrt{-g} \avr{T^{\mu\nu}} \frac{1}{2} \delta g_{\mu\nu}\,,
    \label{eq_delta_W}
\end{align}
thus giving access to the trace $\avr{T^{\mu}_{\ \mu}}$, allowing us to uncover an eventual conformal (trace) anomaly.

The variation of the effective action~\eqref{eq_delta_W} can be expressed in the proper-time representation of Schwinger and DeWitt~\cite{Schwinger1951,DeWitt1975} (in notations of \cite{Brown1977}):
\begin{align}
    \delta W = - \frac{i}{2} \delta {\mathrm{Tr}}\, \int_{0}^\infty \frac{i d s}{i s} e^{- i s H}\,,
    \label{eq_delta_W_2}
\end{align}
via a relativistic Hamiltonian-like operator $H = \Delta + \xi R + m^2$, where a second-order differential operator $\Delta$ represents the kinetic term, the coupling to the curvature $R$ plays the role of an external potential and $m^2$ gives the mass term. The correct analytical properties of Eq.~\eqref{eq_delta_W_2} and similar subsequent relations are maintained by an appropriate complex continuation of the mass term, $m^2 \to m^2 (1 - i 0^+)$, silently assumed here.

The effective Lagrangian~\eqref{eq_W_Leff} takes the following form:
\begin{align}
    L_{\mathrm{eff}} = \frac{1}{2} \frac{1}{(4 \pi)^{\frac{D}{2}}} \int_0^\infty \frac{ids}{(is)^{1+\frac{D}{2}}}
    e^{- i m^2 s} F(x,x;i s;D)\,,
    \label{eq_Leff_ints}
\end{align}
where $F(x,x';i s;D)$ is the weight bi-scalar in the proper-time Green's function $\langle x, s |x',0 \rangle  = \langle x| e^{- i s H} |x' \rangle$ defined in a manner similar to Eq.~\eqref{eq_Leff_ints}. The Green's function satisfies the Schr\"odinger-like equation: $- \frac{\partial }{\partial is} \langle x, s |x',0 \rangle = H \langle x, s |x',0 \rangle$, which gives a quantum-mechanical flavor to the whole proper-time formalism. 

We will not dwell on the precise definition of the bi-scalar $F$, which can be found in detail in Refs.~\cite{DeWitt1975,Brown1977}. The key mathematical point of our arguments is that the function $F$ allows for the power-series expansion in terms of the proper time $s$ (omitting other arguments):
\begin{align}
    F = 1 + i s \, {\mathsf f}_1 + (is)^2 \, {\mathsf f}_2 + \dots\,, 
    \label{eq_F_expansion}
\end{align}
where, in four space-time dimensions, the $O(s^2)$ term captures the trace anomaly~\cite{Brown1977}. On the other hand, the $O(s^2)$ term in an identical\footnote{Taking into account the signs and $i$-th prefactors arising from the difference between Minkowski/Euclidean spacetimes employed in Refs.~\cite{Brown1977,Wondrak2023} one finds that ${\mathsf f}_1 = (\frac{1}{6} - \xi) R$ term in Eq.~(A20) of the proper-time approach of Ref.~[\citeonline{Brown1977}] coincides precisely with the second, $O(s)$ term under the integral in Eq.~(S.17) of the heat-kernel expansion of Ref.~[\citeonline{Wondrak2023}]. Analogously, ${\mathsf f}_2$ in Eq.~(A24) of Ref.~[\citeonline{Brown1977}] coincides precisely with purely gravitational contribution to the third, $O(s^2)$ term under the integral in Eq.~(S.17) of Ref.~[\citeonline{Wondrak2023}]. The ${\mathsf f}_2$ term in series~\eqref{eq_F_expansion} is also reproduced, up to an irrelevant contact term $\Box R$, by the $m=0$ expression in the square brackets of our Eq.~\eqref{eq_A4_explicit} below. Notice that our functions ${\mathsf f}_a$ in Eq.~\eqref{eq_F_expansion} correspond to $f_a$ of Ref.~[\citeonline{Brown1977}] and not to $f_a$ of Ref.~[\citeonline{Wondrak2023}].} expansion of the same 1-loop effective action over the proper time $s$ has been shown in Ref.~[\citeonline{Wondrak2023}] to be associated with the (off-event-horizon) pair-production rate $\Gamma$. The mentioned equivalence of the $O(s^2)$ terms allows us to identify the trace-anomalous origin of the pair production and eventually leads us to Eq.~\eqref{eq_main} as we discuss below.

The renormalized energy-momentum tensor
\begin{align}
    \avr{T^{\mu\nu}}_{\mathrm{ren}} = \frac{1}{4} {\mathcal A}_4 \, g^{\mu\nu} + {\text{non-anomalous part}},
    \label{eq_Tmunu_ren}
\end{align}
contains the anomalous part given by a ${\mathcal A}_4$ function and a non-anomalous part (not shown explicitly). In $D = 4$ spacetime dimensions, the ${\mathcal A}_4$ function in the stress-energy tensor~\eqref{eq_Tmunu_ren} is related to the $O(s^2)$ prefactor in the power series expansion~\eqref{eq_F_expansion} of the bi-scalar $F$~\cite{Brown1977}:
\begin{align}
    {\mathcal A}_4 = \frac{1}{2} \frac{1}{(4 \pi)^2} 
    \left(\frac{\partial }{\partial i s} \right)^2 
    \left[ e^{- i m^2 s} F(x,x;is,4) \right] {\biggl|}_{s = 0}.
\label{eq_A4_d2F}
\end{align}
In the massless theory ($m=0$), the last term in Eq.~\eqref{eq_Tmunu_ren} reduces to a traceless tensor, and the trace of the energy-momentum tensor~\eqref{eq_Tmunu_ren} is fully determined by the trace (scale) anomaly~\eqref{eq_A4_d2F}:
\begin{align}
    \avr{T^\mu_{\ \mu}} \equiv g_{\mu\nu} \avr{T^{\mu\nu}}_{\mathrm{ren}} = {\mathcal A}_4, \qquad\ [\mathrm{for}\ m = 0]\,,
    \label{eq_Tmumu_m0}
\end{align}
where the notation $\avr{T^\mu_{\ \mu}}$ is used for  convenience. Combining Eqs.~\eqref{eq_F_expansion}, \eqref{eq_A4_d2F}, \eqref{eq_Tmumu_m0} and matching them with the $O(s^2)$ term in the effective action of Ref.~[\citeonline{Wondrak2023}] leads us to our main result~\eqref{eq_main}. 

Before proceeding further, we would like to stress that our derivation of Eq.~\eqref{eq_main}, as follows from this section, is applicable only in 3+1 dimensions. One could argue that 1+1 dimensional theories also exhibit Schwinger pair production, with the rate being proportional to the strength of the background electric field~\cite{Dunne:2004nc}. However, this phenomenon corresponds to the time-reversal-odd process, which is associated with the axial anomaly rather than with the conformal anomaly. In simple terms, the axial anomaly means that the axial charge, shown by the axial four-current $J^\mu_A$, is not conserved when there are parallel electric and magnetic fields, which is expressed as $\partial_\mu J^\mu_A \propto \boldsymbol{B} \cdot {\boldsymbol{E}}$. The production of the axial charge is a result of the creation of particles and anti-particles that carry the same axial charge. This process corresponds to both the creation of particles (the Schwinger process) and the breaking of anomalous axial symmetry, which is generally unrelated to the breaking of conformal symmetry. Thus, our Eq.~\eqref{eq_main} is not applicable to the discussed 1+1 dimensional case.

In the rest of the paper, we ensure that Eq.~\eqref{eq_main} is valid for physical environments where both sides of this equation are known. We also discuss photon and neutrino pair production.

\section{Examples}

\subsection{General case} 
A quantum field theory of $N_S$ scalar degrees of freedom, $N_F$ Dirac fermions (a single Majorana or Weyl fermion contributes half of a Dirac fermion, $N_F = 1/2$) and $N_V$ species of massless vector fields, the trace anomaly gets the following form~\cite{
Zeldovich:1977vgo, Starobinskii1981, BD1984, Duff:1993wm, Coriano:2017mux, BS2021}:
\beqn
\avr{T^\mu_{\ \mu}} = b C^2 + b' E_4 + c F_{\mu\nu} F^{\mu\nu}\,, \qquad
\label{eq_Tmunu_general}
\eeqn
where $C^2 = R_{\mu\nu\alpha\beta} R^{\mu\nu\alpha\beta} - 2 R_{\mu\nu} R^{\mu\nu} + R^2/3$
is the Weyl tensor squared and $E_4 = R_{\mu\nu\alpha\beta} R^{\mu\nu\alpha\beta} - 4 R_{\mu\nu} R^{\mu\nu} + R^2$ is the Euler density in $D=4$ dimensions. In Eq.~\eqref{eq_Tmunu_general}, the physically irrelevant $\Box R$ term is omitted and the conformal coupling ($\xi = 1/6$ for scalars) is assumed. The parameters are as follows~\cite{Duff:1993wm,Capper:1973mv,Capper:1974ed,Capper:1974ic,Brown1977,Christensen:1977jc}:
\begin{align}
    b  & = \phantom{-} \frac{1}{120} \frac{1}{(4\pi)^2} \left(N_S + 6 N_F + 12 N_V\right)\,, \\
    b' & = - \frac{1}{360} \frac{1}{(4\pi)^2} \left(N_S + 11 N_F + 62 N_V\right)\,.
\end{align}
As a check, one finds that in a pure gravitational background ($F_{\mu\nu} = 0$), $N_S = N$ species of neutral scalar fields (with $N_F = N_V = 0$), Eq.~\eqref{eq_Tmunu_general} reduces to Eq.~\eqref{eq_A4_explicit} with a factor $N$ and leads us, via Eq.~\eqref{eq_main}, to Eq.~\eqref{eq_PRL_recent} as expected.

The last term~\eqref{eq_Tmunu_general} represents a non-universal (``matter'') part which accounts for renormalization effects related to the scale dependence of the couplings of the theory with a vector-field background. It depends only on the {\it first} electromagnetic invariant, $F_{\mu\nu} F^{\mu\nu}$. A nontrivial scalar background in an interacting scalar field theory can also generate a matter-type contribution~\cite{vanNieuwenhuizen:1999nu,Nojiri:1997xk,Asorey:2022ebz}.

For gauge vector fields coupled minimally with matter fields via the electric coupling~$e$, the prefactor\footnote{Notice the minus sign in Eq.~\eqref{eq_c} corresponding to the mostly-plus metric convention adopted in this article ({\it cf.}, for example, Ref.~[\citeonline{Giannotti2009}]).} 
\begin{align}
    c = - \frac{\beta(e)}{2 e}\,,
\label{eq_c}
\end{align}
of the last term in Eq.~\eqref{eq_Tmunu_general} depends on the beta function $\beta(e) = \mu \, {\mathrm d} e/{\mathrm d} \mu$ associated with the running of the coupling $e$. A nonvanishing beta function expresses the fact that radiative corrections make the electric charge $e = e(\mu)$ sensitive to the renormalization energy scale~$\mu$, thus breaking the scale invariance of the system.

The gravitational part of the anomaly represented by the first two terms in Eq.~\eqref{eq_Tmunu_general} is exact in one loop, as higher-order corrections to this expression vanish. This statement is not true for the third (matter) term since radiative corrections exist, generally, in all loops~\cite{Shifman:1988zk}.

\subsection{A neutral scalar field in curved spacetime}

As the first check of our formula~\eqref{eq_main}, we consider a single-component neutral scalar field of mass $m$ and non-con\-formal coupling $\xi$ to gravity described by Lagrangian~\eqref{eq_L_scalar}. It is well known that quantum fluctuations in this theory produce the following trace anomaly~\cite{Deser1976,Brown1977}:
\begin{align} \label{eq_A4_explicit}
    \avr{T^\mu_{\ \mu}}_{1\mathrm{sc}} 
    = & \frac{1}{(4\pi)^2}\biggl[ \frac{1}{180} R_{\mu\nu\alpha\beta} R^{\mu\nu\alpha\beta} 
    - \frac{1}{180}  R_{\mu\nu} R^{\mu\nu} \\
    & + \frac{1}{6} \left(\frac{1}{5} - \xi \right) \Box R + \frac{1}{2} \left(\frac{1}{6} - \xi \right)^2 R^2 + \frac{1}{2} m^4 \biggr]\,.
    \nonumber 
\end{align}
Combining Eqs.~\eqref{eq_A4_explicit} and \eqref{eq_main}, we recover exactly the result of Ref.~[\citeonline{Wondrak2023}] given in Eq.~\eqref{eq_PRL_recent} for a single ($N=1$) neutral ($q=0$) massless ($m=0$) scalar field. Notice that the $\Box R$ term, present in Eq.~\eqref{eq_A4_explicit} and absent in Eq.~\eqref{eq_PRL_recent}, can be removed by a finite local counterterm during the renormalization procedure and hence is physically irrelevant.  

\subsection{Scalar QED in flat spacetime} A similarity between the gravitational particle production and the Schwinger pair production in flat spacetime in the background electric field has been noticed in Ref.~[\citeonline{Wondrak2023}] on the basis of equation~\eqref{eq_PRL_recent}. Here we show that the conformal anomaly plays the crucial role in this relation.

Consider a theory of $N_S$ species of massless complex scalars coupled to electromagnetism with the same electric coupling $e$ (a ``scalar Quantum Electrodynamics'' or sQED) in a flat spacetime where the model possesses the radiative contribution to the trace anomaly due to non-zero beta function~\cite{Weisskopf:1936hya,Dunne:2004nc},
\beqn
\beta_{{\text{sQED}}}^{\mathrm{1loop}}  = \frac{N_S e^3}{48 \pi^2}\,,
\label{eq_beta_sQED}
\eeqn
given here in one loop. Equations~\eqref{eq_c} and \eqref{eq_beta_sQED} imply that the coefficient in the last term of the trace anomaly~\eqref{eq_Tmunu_general} is $c = N_S e^2/(96 \pi^2)$. Then Eq.~\eqref{eq_main} gives us $\Gamma_{\mathrm{sQED}} = - N_S e^2 F_{\mu\nu} F^{\mu\nu}/(192 \pi)$, which exactly coincides with the pair production rate~\eqref{eq_PRL_recent} of Ref.~[\citeonline{Wondrak2023}] if one takes into account that each complex field carries two degrees of freedom: $N = 2 N_S$.

According to our convention in Eq.~\eqref{eq_main}, there is no particle production for a negative production rate. Since $F_{\mu\nu} F^{\mu\nu} = 2 ({\bs B}^2 - {\bs E}^2)$, Eq.~\eqref{eq_main} implies the absence of particle creation in a pure magnetic field in scalar QED because $\Gamma_{\mathrm{sQED}} < 0$. However, in the electric-field background, one gets the following well-known result for the complex scalar field (reproduced also in Ref.~[\citeonline{Wondrak2023}] for $N_S = 1$):
\begin{align}
    \Gamma_{\mathrm{sQED}} = N_S \frac{e^2 {\bs E}^2}{96 \pi}\,.
\label{eq_Gamma_sQED}
\end{align}
This equivalence further supports the validity of Eq.~\eqref{eq_main}.

\subsection{Spinor QED in flat spacetime} 

Since the one-loop beta function of the massless spinor QED with $N_F$ flavors of massless Dirac fermions is four times bigger (per particle) than its scalar analogue~\eqref{eq_beta_QED}~\cite{Dunne:2004nc}:
\beqn
\beta_{{\text{QED}}}^{\mathrm{1loop}}  = \frac{e^3}{12 \pi^2}\,,
\label{eq_beta_QED}
\eeqn
we get that the pair particle production rate reduces exactly to the well-known massless QED result~\cite{Schwinger1951,Dunne:2004nc}:
\begin{align}
    \Gamma^{(m=0)}_{\mathrm{QED}} = N_F \frac{e^2 {\bs E}^2}{24 \pi} \qquad \mathrm{[electromagnetic]}\,.
\label{eq_Gamma_QED_flat}
\end{align}
Thus, Eq.~\eqref{eq_main} matches the pair production rate~\eqref{eq_Gamma_QED_flat} and the one-loop trace anomaly in spinor QED, $\langle T^\mu_{\mu}\rangle  = \frac{e^2}{24 \pi^2} F_{\mu\nu} F^{\mu\nu}$~[\citeonline{Crewther1972, Chanowitz1972, Chanowitz1973a, Chanowitz1973b, Adler1977, Drummond1980}].

The proportionality of the pair-creation rates for scalar \eqref{eq_Gamma_sQED} and spinor~\eqref{eq_Gamma_QED_flat} QED to their beta functions, Eqs.~\eqref{eq_beta_sQED} and \eqref{eq_beta_QED}, respectively, is not surprising given an intimate relation between the effective Euler-Heisenberg Lagrangian and the beta function (for an excellent review, see Ref.~[\citeonline{Dunne:2004nc}]). As the beta function also contributes to the trace anomaly, the relation of the trace anomaly to the pair-creation rate closes the logical triangle, thus qualitatively supporting Eq.~\eqref{eq_main} on physical grounds.

Relation~\eqref{eq_main} suggests that the creation rate of pairs of massless particles in flat spacetime in a classical electromagnetic background is related to the beta function:
\begin{align}
    \Gamma_{\mathrm{flat}} = - \frac{\pi \beta(e)}{4e} F_{\mu\nu} F^{\mu\nu}\,.
    \label{eq_Gamma_generic}
\end{align}
Equation~\eqref{eq_Gamma_generic} is valid at least in one-loop order with the mentioned reservation that a negative production rate implies the absence of pair production. 

The rate~\eqref{eq_Gamma_generic} takes into account {\it only} the conformal contribution, which depends on the $T$-even electromagnetic invariant $F_{\mu\nu} F^{\mu\nu}$. It does not account for the axial anomaly, which appears, for example, in parallel electric and magnetic fields ${\bs E} \| {\bs B}$. The corresponding contribution to the pair production depends on the $T$-odd electromagnetic invariant $\epsilon^{\mu \nu \rho \sigma} F_{\mu\nu} F_{\rho\sigma}$, giving $\Gamma^{\mathrm{axial}} \sim E B$ for massless Dirac fermions~\cite{Dunne:2004nc} (see also Comment~[\citeonline{Ferreiro2024}] on Ref.~[\citeonline{Wondrak2023}]). Likewise, the mixed axial-gravitational anomaly~\cite{AlvarezGaum1984}, set by the $T$-odd invariant $\epsilon^{\mu\nu \rho\sigma} R^{\alpha}_{\ \beta\mu\nu} R^{\beta}_{\ \alpha\rho\sigma}$, does not enter our main conformal result~\eqref{eq_main}.

One should notice that the particle creation rates in both scalar~\eqref{eq_Gamma_sQED} and spinor~\eqref{eq_Gamma_QED_flat} QED should be interpreted under the assumption that the backreaction and screening of the electric field by the produced pairs are neglected. Indeed, the absence of an energy gap in the massless theory leads to an infinite density of low-energy states, allowing for unbounded production of soft pairs in a constant background electric field. These processes lead to the infrared divergence of the Schwinger pair production~\cite{Dunne:2004nc}. The same cautionary assumption is applied to our result~\eqref{eq_main}.

\subsection{Schwarzschild spacetime} 

Consider now a purely gravitational background given by the static Schwarzschild spacetime of a body with the mass $M$:
\begin{align}
    d s^2 = &\, - \left( 1 - \frac{2 M G}{r} \right) dt^2 + \left( 1 - \frac{2 M G}{r} \right)^{-1} d r^2 + r^2 (d \theta^2 + \sin^2\theta d \varphi^2)\,.
\end{align}
Given Ricci flatness ($R_{\mu\nu} = 0$) of this metric, the gravitational contribution to the pair-creation rate is provided only by the Kretschmann scalar: $K = R_{\mu\nu\alpha\beta} R^{\mu\nu\alpha\beta} = 48 G^2 M^2/r^6 > 0$, with the result~\eqref{eq_main}:
\begin{align}
    \Gamma_{\mathrm{grav}} & = 
     \left(2 N_S + 7 N_F - 26 N_V\right) \frac{1}{240 \pi} \frac{G^2 M^2}{r^6}\,,
\label{eq_Gamma_grav}
\end{align}
for a free field theory with scalar, spinor and vector massless fields. Equation~\eqref{eq_Gamma_grav}, gives us {\it partial} production rates for corresponding species. Notice that the pair production rate~\eqref{eq_Gamma_grav} is exactly proportional to the coefficient of the logarithmic terms for the entanglement entropy of the quantum fields for Schwarzschild black hole~\cite{Solodukhin:1994yz,Fursaev:1994te,Solodukhin:1997yy,Solodukhin:2019xwx} suggesting an existence of a relation between them.

In realistic QED in weak electromagnetic fields, the four-photon scattering can be neglected~\cite{Dunne:2004nc}, so that photon propagation can be described by free Maxwell theory with simple Lagrangian, ${\mathcal L}_{\mathrm{ph}} = - (1/4) F_{\mu\nu} F^{\mu\nu}$. The particle creation rate corresponds to $N_F = 0$, $N_S = 0$ and $N_V = 1$ in Eq.~\eqref{eq_Gamma_grav} and gives us the following discouraging result $\Gamma_{\mathrm{ph}} = - 13 G^2 M^2/(120 \pi r^4) < 0$ implying that no photons can be created in the gravitational field due to the conformal anomaly mechanism.

Similar considerations can also be applied to the neutrino--anti-neutrino pair creation with the appropriate replacement of the spinor degrees of freedom by the sum of Dirac and Majorana neutrino species: $N_\nu \to N_\nu = N_D + (1/2) N_M$ and taking $N_V = N_S = 0$ in Eq.~\eqref{eq_Gamma_grav}. One gets the neutrino pair production $\Gamma_{\nu} = 7 N_\nu G^2 M^2/(240 \pi r^4) > 0$, which is $7/2$ times bigger than the rate of pair production for scalars (the estimations of the latter can be found in Ref.~[\citeonline{Wondrak2023}]).

One could ask whether these results are applicable to massive particles. For the pure electromagnetic contribution to the pair creation rate, the condition is well known~\cite{Dunne:2004nc}: the electric field strength should substantially exceed the critical Schwinger field, $E \gg E_c^{\mathrm{Sch}} = m^2_e/e \simeq 1.3\times 10^{18}\, \mathrm{V/m}$. Likewise, the same condition can be obtained by demanding that the gravitational contribution should exceed the anticipated $\propto m_e^4$ term generated by the explicit breaking of the conformal symmetry [{\it cf.} Eq.~\eqref{eq_A4_explicit}]. For neutrinos, the applicability condition reads as $r \ll r_c$ with the critical radius $r_c = \sqrt{GM}/m_\nu$.

Following Ref.~[\citeonline{Duff:1993wm}], we suggest that in conformally broken theories ({\it e.g.}, for massive fields), the right-hand side of Eq.~\eqref{eq_main} should be modified: $\avr{T^\mu_{\ \mu}} \to g_{\mu\nu}\avr{T^{\mu\nu}}_{\mathrm{ren}} - \avr{g_{\mu\nu} T^{\mu\nu}}_{\mathrm{ren}}$, implying  that the explicitly non-conformal mass term $m^4$ does not contribute to~\eqref{eq_main}.

\subsection{Savvidi magnetic instability in QCD}

In the pure magnetic field, both in scalar QED and spinor QED, the right-hand side of Eq.~\eqref{eq_main} is a negative quantity and, therefore, no instability associated with the particle production can occur. Of course, this natural conclusion is supported by the fact that their beta functions, Eqs.~\eqref{eq_beta_sQED} and \eqref{eq_beta_QED}, are positively defined. But what happens if the beta function is negative? 

Consider, for example, Yang-Mills (YM) theory, which determines non-perturbative properties of Quantum Chromodynamics (QCD). The beta function of $N_c$-color YM theory, $\beta_{\mathrm{YM}}(g) = - 11 N_c g^3/(48 \pi^2)$, is a negative function of the strong coupling constant $g$. Adopting  Eq.~\eqref{eq_Gamma_generic} to non-Abelian fields possessing the field strengths $F_{\mu\nu}^a$, one gets the following {\it {formal perturbative expression}} for the gluon production rate:
\begin{align}
    \Gamma^{\mathrm{pert}}_{\mathrm{YM}} & 
    = \frac{11 N_c g^2}{192 \pi} F^a_{\mu\nu} F^{a,\mu\nu} 
    \equiv \frac{11 N_c}{96 \pi} \left[ (g {\bs B}^a)^2 - (g {\bs E}^a)^2 \right] \,,
\label{eq_Gamma_YM}
\end{align}
where the sum over gluons, $a = 1, \dots, N_c^2 -1$, is assumed. 

Equation~\eqref{eq_Gamma_YM} represents a formal expression that is not applicable to the ground state of YM theory because Eq.~\eqref{eq_Gamma_YM} corresponds to the anomalous breaking of scale symmetry associated with the perturbative renormalization of couplings -- hence the superscript ``pert'' in Eq.~\eqref{eq_Gamma_YM} --  while in YM theory, the conformal symmetry is broken dynamically and non-perturbatively~\footnote{In QCD, the magnitude of the  dynamical breaking of conformal symmetry, $\avr{T^\mu_{\ \mu}} \simeq \Lambda_{\mathrm{QCD}}^4$, is determined by an intrinsic mass scale $\Lambda_{\mathrm{QCD}}$ of the order of a few hundred MeV.}. 
Despite this fact, Eq.~\eqref{eq_Gamma_YM} still allows us to make another interesting relationship with an already known effect: the instability of the perturbative gluonic vacuum. Indeed, since $\Gamma_{\mathrm{YM}} = 11 N_c (g {\bs B}^a)^2/(96 \pi) > 0$, even the weakest background gluomagnetic field leads to the creation of gluon pairs and makes the gluonic vacuum unstable. This observation matches well with the instability of the perturbative gluon vacuum~\cite{Nielsen:1978rm} which drives creation of the magnetic condensate (the Savvidi vacuum~\cite{Savvidy:1977as}) and the formation of the suggested magnetic-spaghetti vacuum state~\cite{Nielsen:1979xu,Ambjorn:1979xi} precisely due to the negativeness of the YM beta function $\beta_{\mathrm{YM}}(g) < 0$ (see also \cite{Yildiz1980}).

\section{Critical Remarks} Our formula~\eqref{eq_main}, which reproduces the main result of Ref.~[\citeonline{Wondrak2023}], recovers the correct expression for the pair production of massless particles in a background electric field in the Abelian gauge theories (in which the beta function is positive) and aligns well with the vacuum instability in a background magnetic field in the non-Abelian Yang-Mills theories (where the beta function is negative). However, the genuine picture of the advocated conformal mechanism of the particle creation in the static gravitational background is not clear because of the absence of a uniquely defined vacuum state in a general curved spacetime~\cite{Fulling1973, Kay1991}. In contrast, our main result~\eqref{eq_main} is formulated independently of any particular choice of quantum state, which may imply that this formula -- if applied to a gravitational background -- works only for a particular choice of the vacuum state. Moreover, one should mention that the gravitational part of the particle creation mechanism, suggested in Ref.~[\citeonline{Wondrak2023}], remains under debate, as discussed in Ref.~[\citeonline{Ferreiro2024, Hertzberg:2023xve, Akhmedov2024}], to which we refer an interested reader.

\section{Conclusions}

We suggested the simple formula~\eqref{eq_main} for the $T$-even contribution to the particle production rate. We argued that the underlying mechanism of the particle production is related to the anomalous breaking of conformal symmetry. Equation~\eqref{eq_main} reproduces known results for the Schwinger pair production both in scalar and spinor QED and correctly indicates instability of the perturbative QCD vacuum.  It also reproduces the expression for the particle production in curved spacetime and in electromagnetic background fields proposed in Ref.~[\citeonline{Wondrak2023}]. The conformal part of the particle production does not include the effect from the axial anomaly, which depends on $T$-odd invariants that are pertinent, for example, to parallel electric and magnetic fields. A word of caution is also given for the ambiguity related to different vacua in the case of gravitational pair production.

\section*{Acknowledgments}
The correspondence with A. D.~Dolgov, S. A. Franchino-Vi\~nas, I. L.~Shapiro, S. N. Solodukhin and G. E.~Volovik is gratefully acknowledged. This work has been supported by the French National Agency for Research (ANR) within the project PROCURPHY ANR-23-CE30-0051-02.

\end{document}